\documentclass[twocolumn]{aastex631}

\usepackage{xspace}
\usepackage{CJK}
\usepackage{rotating}
\usepackage{afterpage}
\usepackage{changepage}
\usepackage{sidecap}
\usepackage{comment}

\hypersetup{
    colorlinks=true,
    linkcolor=blue,
    filecolor=magenta,      
    urlcolor=cyan,
}

\usepackage[normalem]{ulem} % for regular text; use \sout{}
\usepackage{cancel} % for equations; use \cancel{}

\usepackage{color}

\newcommand{\kms}{\ensuremath{\rm{km\,s}^{-1}}\xspace}
\newcommand{\alphaco}{\ensuremath{\alpha_{\rm{CO}}}\xspace}

\newcommand{\uJybeam}{\ensuremath{\mu\rm{Jy}\,\rm{beam}^{-1}}\xspace}

\newcommand{\Mstar}{\ensuremath{M_{\rm{*}}}\xspace}

\newcommand{\MHt}{\ensuremath{M_{\rm{H}_2}}\xspace}

\newcommand{\Msol}{\ensuremath{M_\odot}\xspace}
\newcommand{\sfr}{\ensuremath{\rm{M}_\odot\,yr^{-1}}\xspace}

\newcommand{\etal}{et~al.\xspace}

\newcommand{\arc}{\ensuremath{''}\xspace}
\newcommand{\squiggle}{SQuIGG\ensuremath{\vec{L}}E\xspace}

\newcommand{\ratio}{\ensuremath{L'_{\rm{CO(5-4)}}/L'_{\rm{CO(2-1)}}}\xspace}

\shortauthors{D'Onofrio, et~al.}
\shorttitle{Gas Excitation in \squiggle PSBs}

\begin{document}
\begin{CJK*}{UTF8}{gbsn}

\title{Molecular Gas Excitation in $z\sim0.7$ Gas-Rich Post-starburst Galaxies from \squiggle}

\correspondingauthor{Vincenzo~R.~D'Onofrio}
\email{donofr19@tamu.edu}

\author[0000-0002-1759-6205]{Vincenzo~R.~D'Onofrio}
\affiliation{Department of Physics and Astronomy and George P. and Cynthia Woods Mitchell Institute for Fundamental Physics and Astronomy, Texas A\&M University, 4242 TAMU, College Station, TX 77843-4242, US}

\author[0000-0003-3256-5615]{Justin~S.~Spilker}
\affiliation{Department of Physics and Astronomy and George P. and Cynthia Woods Mitchell Institute for Fundamental Physics and Astronomy, Texas A\&M University, 4242 TAMU, College Station, TX 77843-4242, US}

\author[0000-0001-5063-8254]{Rachel~Bezanson}
\affiliation{Department of Physics and Astronomy and PITT PACC, University of Pittsburgh, Pittsburgh, PA 15260, USA}

\author[0000-0002-1109-1919]{Robert~Feldmann}
\affiliation{Department of Astrophysics, University of Zurich, Winterthurerstrasse 190, Zurich CH-8057, Switzerland}

\author[0000-0003-4700-663X]{Andy~D.~Goulding}
\affiliation{Department of Astrophysical Sciences, Princeton University, Princeton, NJ 08544, USA}

\author[0000-0002-5612-3427]{Jenny~E.~Greene}
\affiliation{Department of Astrophysical Sciences, Princeton University, Princeton, NJ 08544, USA}

\author[0000-0002-7613-9872]{Mariska~Kriek}
\affiliation{Leiden Observatory, Leiden University, P.O. Box 9513, 2300 RA Leiden, The Netherlands}

\author[0009-0005-4226-0964]{Anika~Kumar} 
\affiliation{Laboratory for Multiwavelength Astrophysics, School of Physics and Astronomy, Rochester Institute of Technology, 84 Lomb Memorial Drive, Rochester, NY 14623, USA} 
\affiliation{Department of Physics and Astronomy and PITT PACC, University of Pittsburgh, Pittsburgh, PA 15260, USA} 

\author[0000-0002-0696-6952]{Yuanze~Luo}
\affiliation{Department of Physics and Astronomy and George P. and Cynthia Woods Mitchell Institute for Fundamental Physics and Astronomy, Texas A\&M University, 4242 TAMU, College Station, TX 77843-4242, US}

\author[0000-0002-7064-4309]{Desika~Narayanan}
\affiliation{Department of Astronomy, University of Florida, 211 Bryant Space Science Center, Gainesville, FL 32611, USA}

\author[0000-0003-4075-7393]{David~J.~Setton}\thanks{Brinson Prize Fellow}
\affiliation{Department of Astrophysical Sciences, Princeton University, Princeton, NJ 08544, USA}

\author[0000-0002-1714-1905]{Katherine~A.~Suess}
\affiliation{Department for Astrophysical \& Planetary Science, University of Colorado, Boulder, CO 80309, USA}
%\altaffiliation{NHFP Hubble Fellow}
%\affiliation{Kavli Institute for Particle Astrophysics and Cosmology and Department of Physics, Stanford University, Stanford, CA 94305, USA.}

\author[0000-0003-1535-4277]{Margaret~E.~Verrico}
\affiliation{University of Illinois Urbana-Champaign Department of Astronomy, University of Illinois, 1002 W. Green St., Urbana, IL 61801, USA}
\affiliation{Center for AstroPhysical Surveys, National Center for Supercomputing Applications, 1205 West Clark Street, Urbana, IL 61801, USA}

%%%%%%%%%%%%%%%%%%%%%%%%%%%%%%%%%%%%%%%%%%%%%%%%%%%%%%%%%%%%%%%%%%%%%%%%%%%%%%%%%%%%%
%%%%%%%%%%%%%%%%%%%%%%%%%%%%%%%%%%%% ABSTRACT %%%%%%%%%%%%%%%%%%%%%%%%%%%%%%%%%%%%%%%
%%%%%%%%%%%%%%%%%%%%%%%%%%%%%%%%%%%%%%%%%%%%%%%%%%%%%%%%%%%%%%%%%%%%%%%%%%%%%%%%%%%%%
\begin{abstract}
\noindent Many post-starburst galaxies at $z\sim0.7$ have been shown to retain substantial molecular gas reservoirs yet host low ongoing star formation, suggesting that the remaining gas may be inefficient at forming stars during the early post-burst phase. We present new Atacama Large Millimeter/submillimeter Array CO(5--4) observations of nine gas-rich post-starburst galaxies at $z\sim0.7$ from the Studying Quenching in Intermediate-z Galaxies: Gas, angu$\vec{L}$ar momentum, and Evolution (\squiggle) survey, providing a view of the molecular gas excitation in these systems. Combined with existing CO(2--1) data, we detect CO(5--4) in 8/9 targets and find that most have moderate CO excitation with $r_{52}\equiv L'_{\rm CO(5-4)}/L'_{\rm CO(2-1)}\approx0.1-0.3$. These systems show no clear trend between $r_{52}$ and either total or surface-density of star formation. Specifically, all objects have $\Sigma_{\rm SFR}\sim0.01-1$\,\sfr\,kpc$^{-2}$, consistent with compact, modest star formation, even when allowing for buried activity, as these galaxies decline from their peak. One object J1448+1010, which has clear optical, mid-infrared, and radio indicators of an active galactic nucleus, is an outlier with $r_{52}\approx0.6$; its elevated excitation likely requires significant non-stellar heating, with a contribution from potentially obscured star formation. Together, most gas-rich \squiggle post-starbursts have moderately excited molecular gas alongside little to modest star-forming activity, indicating that the remaining gas hosts relatively suppressed star formation efficiencies instead of strong buried starburst activity. 
\end{abstract}

%%%%%%%%%%%%%%%%%%%%%%%%%%%%%%%%%%%%%%%%%%%%%%%%%%%%%%%%%%%%%%%%%%%%%%%%%%%%%%%%%%%%%
%%%%%%%%%%%%%%%%%%%%%%%%%%%%%%%%%% Introduction %%%%%%%%%%%%%%%%%%%%%%%%%%%%%%%%%%%%%
%%%%%%%%%%%%%%%%%%%%%%%%%%%%%%%%%%%%%%%%%%%%%%%%%%%%%%%%%%%%%%%%%%%%%%%%%%%%%%%%%%%%%
\section{Introduction} \label{intro}

The rapid cessation of star formation in massive galaxies remains a fundamental problem in galaxy evolution. Although large-scale surveys have firmly established a bimodality in galaxy populations (e.g., \citealt{kauffmann_dependence_2003}), the physical mechanisms responsible for transforming star-forming galaxies into quiescent systems remain debated. Many quenching models invoke processes that affect the molecular gas, typically by heating or expelling the reservoirs (e.g., \citealt{di_matteo_energy_2005, hopkins_unified_2006, croton_many_2006}). Likewise, observational evidence for outflows driven by active galactic nuclei (AGN), merger-induced shocks, and environmental interactions demonstrate that feedback can strongly perturb the molecular interstellar medium (ISM) and suppress its ability to form stars (e.g., \citealt{alatalo_escape_2015, fluetsch_cold_2019, donofrio_quenching_2025}). However, simply the presence or absence of molecular gas is not sufficient to distinguish among these scenarios. Instead the physical state of the gas, set primarily by its density and kinetic temperature, traces how feedback has acted on the ISM (e.g., \citealt{weis_highly-excited_2007,narayanan_theory_2014,kamenetzky_recovering_2018}).

Post-starburst galaxies (PSBs) offer a unique view to investigate how the physical state of the gas evolves during quenching. Their A-star-dominated optical spectra, with strong Balmer absorption and weak nebular emission, indicate a recent burst of star formation followed by a rapid decline within the past $\sim$0.1-1\,Gyr (see \citealt{french_evolution_2021} for a review of PSBs). While rare at all epochs, by $z\sim2$ PSBs comprise $\sim$5\% of massive galaxies and account for nearly half of the new additions to the massive red sequence (e.g., \citealt{wild_evolution_2016,belli_mosfire_2019,park_rapid_2023}). Observations of low-redshift PSBs have revealed that many retain substantial molecular gas reservoirs post-quenching, with gas fractions comparable to star-forming galaxies of similar mass \citep{french_discovery_2015,rowlands_evolution_2015,alatalo_shocked_2016}. 

The CO excitation provides a direct probe of the physical state of the molecular gas, where the relative strengths of different transitions trace its characteristic temperature and density. Although these measurements do not extend to the higher-$J$ regime, CO lines up to $J=3$ together with dense gas tracers (e.g., HCN, HCO$^+$) in low-redshift PSBs indicate that the remaining gas is predominantly diffuse and inefficient at forming stars, likely to fade over long depletion times without substantial regrowth \citep{french_state_2023}. In this sense their CO ladders resemble those of quiescent systems, with low dense gas fractions. However, a contrasting example is the local shocked PSB NGC 1266 (with CO ladder extending up to $J=13$) which exhibits markedly elevated CO excitation in the dense $\lesssim$100\,pc nucleus \citep{alatalo_discovery_2011,pellegrini_shock_2013}. This excitation is likely driven by a combination of low-level compact star formation and mechanical or radiative heating associated with an AGN, where outflow-driven shocks inject turbulence that suppresses star formation outside the nucleus and gradually depletes the extended diffuse gas reservoir   \citep{pellegrini_shock_2013,alatalo_escape_2015,otter_pulling_2024,otter_clumpy_2026}.  

At higher redshift, where galaxies are richer in gas (e.g., \citealt{tacconi_evolution_2020}) and mergers are more common (e.g., \citealt{lotz_evolution_2008,conselice_structures_2009,bridge_cfhtls-deep_2010,duncan_observational_2019,ferreira_galaxy_2020}), it remains unclear which excitation regime PSBs occupy since the post-burst molecular ISM conditions may differ at these epochs. Recent observations of $z\sim0.6-1.3$ PSBs reveal on average modest CO excitation ($\langle r_{52}\rangle \sim 0.3$ in brightness temperature units) from mostly nondetections with some variation among systems \citep{zanella_gas_2025}, highlighting the need for a uniformly selected sample at this epoch to establish how the ISM evolves during this phase.     
The Studying Quenching in Intermediate-z Galaxies: Gas, angu$\vec{L}$ar momentum, and Evolution (\squiggle) program \citep{suess_mathrmsquiggecle_2022} selected 1318 massive PSBs at $z\sim0.7$ from the Sloan Digital Sky Survey (SDSS) DR14 spectroscopic database \citep{abolfathi_fourteenth_2018}, designed to select galaxies that have just ended their primary epoch of star formation. Atacama Large Millimeter/submillimeter Array (ALMA) CO(2--1) observations of a representative subset of 50 galaxies reveal that, similar to low-redshift PSBs, quenching can precede the removal of cold gas in higher-redshift systems, with large H$_2$ reservoirs persisting immediately after the starburst and then fading on $\sim$100-200\,Myr timescales \citep{suess_massive_2017,bezanson_now_2022,setton_squigg_2025}. In multiple cases, up to roughly half of the molecular gas in the system resides in extended tidal tails as long as $\sim$60\,kpc, while the gas remaining in the central galaxies is highly turbulent and inefficient at forming stars \citep{spilker_star_2022,donofrio_quenching_2025}. These results highlight the importance of directly measuring the CO excitation of the molecular gas in gas-rich \squiggle PSBs to test why these reservoirs are not actively forming stars.

An additional possibility is that a subset of the \squiggle galaxies may host significant dust-obscured star formation that was not captured by the optical selection. Several systems exhibit very large inferred molecular gas reservoirs relative to their low apparent star formation rates (SFRs), raising the possibility that some objects classified as PSB instead may be actively forming stars but are heavily obscured, as suggested by simulations of $z\sim0.7$ PSBs \citep{cenci_nature_2025}. This scenario was explored empirically by \citet{setton_squigg_2025}, who showed that the intrinsic SFRs of \squiggle PSBs could be $\sim$0.5 dex higher when incorporating mid- and far-infrared photometry, although their analysis found no strong preference for this interpretation over models invoking genuinely suppressed star formation. In the case that substantial obscured star formation is present and directly coupled to the molecular gas, the density and temperature of the gas would be raised, and potentially accompanied by substantial mid-$J$ CO emission.  

\begin{figure*}[ht!]
\centering
\hspace{0.0cm}
\includegraphics[width=0.9\textwidth]{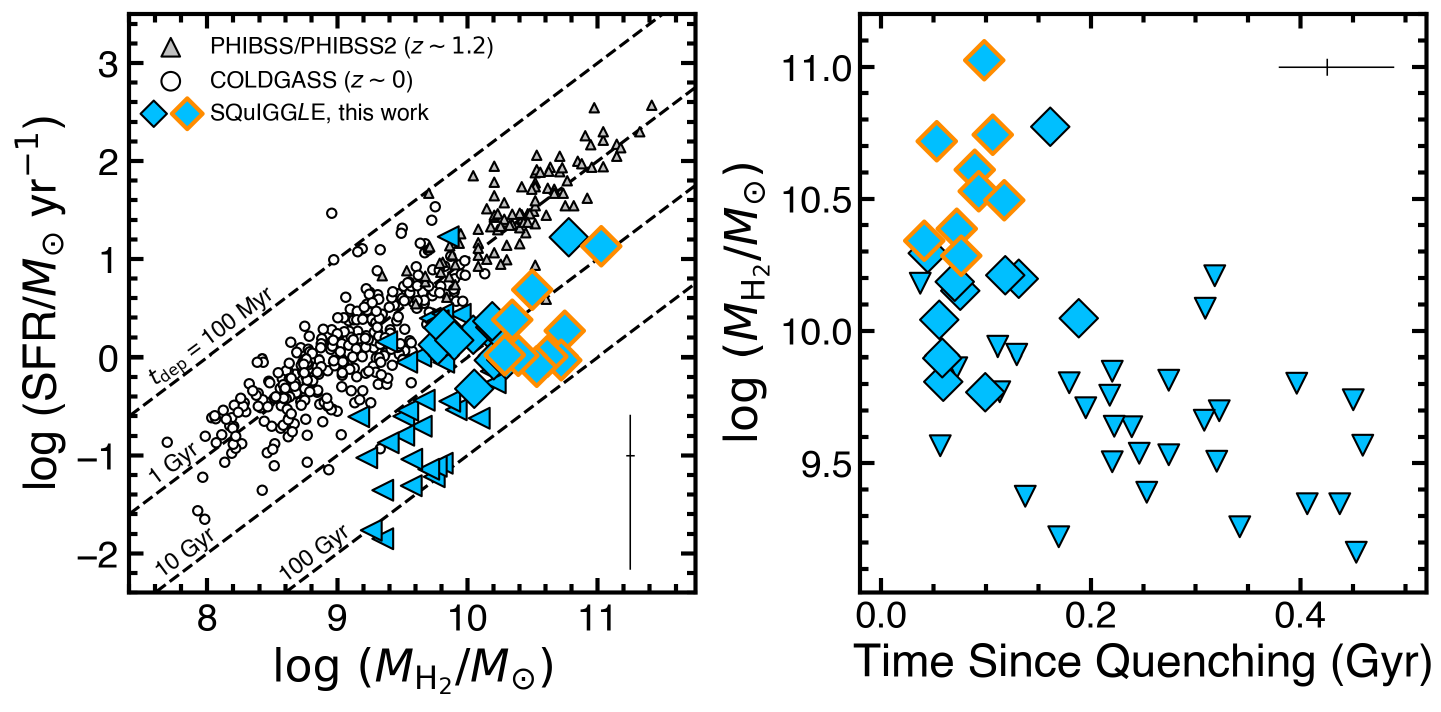}
\caption{SFR vs. H$_2$ gas mass (left) for all CO(2--1) \squiggle measurements (blue diamonds) and nondetections (blue triangles), with the nine PSBs targeted in this work outlined in orange. We include for comparison star-forming galaxies at $\langle z \rangle \sim1.2$ from PHIBSS/PHIBSS2 (gray triangles; \citealt{tacconi_phibss_2013}) along with $z\sim0$ massive galaxies from COLDGASS (white circles; \citealt{saintonge_cold_2011}). Black dashed lines indicate constant molecular gas depletion timescales ($\tau_\mathrm{dep} \equiv {M}_{\mathrm{H_2}}/\mathrm{SFR}$). \MHt vs. time since quenching $t_q$ (right) for the same sample. H$_2$ masses are taken from \citet{bezanson_now_2022} and \citep{setton_squigg_2025}, while SFR and $t_q$ values are from \citet{suess_mathrmsquiggecle_2022}. Median uncertainties for the entire \squiggle ALMA sample are shown for both panels (black cross). The selected systems are among the most gas-rich and youngest PSBs in the \squiggle sample.
}
\label{fig:MgasSFR}
\end{figure*}

In this work, we use new ALMA Band 7 observations of the CO(5--4) transition to probe the excitation of the molecular gas in nine gas-rich \squiggle PSBs with existing CO(2--1) measurements. Together, these data allow us to assess whether gas-rich \squiggle PSBs host molecular gas with CO excitation consistent with diffuse and quiescent systems similar to the physical state of the remaining gas reservoirs in most low-redshift PSBs, or whether their CO excitation suggests a warm and dense origin that may imply ongoing star formation. In Section~\ref{data} we describe the sample and ALMA observations. In Section~\ref{Results} we present the CO(5--4) measurements and examine the relationship between CO excitation and potential heating origins. Finally, we summarize our conclusions in Section~\ref{Conclusions}. We assume throughout this work a \citet{chabrier_galactic_2003} initial mass function (IMF) and a concordance Lambda cold dark matter cosmology with $\Omega_m=0.3$ and $h=0.7$.

%%%%%%%%%%%%%%%%%%%%%%%%%%%%%%%%%%%%%%%%%%%%%%%%%%%%%%%%%%%%%%%%%%%%%%%%%%%%%%%%%%%%%
%%%%%%%%%%%%%%%%%%%%%%%%%%%%%%%%% Data & Methods %%%%%%%%%%%%%%%%%%%%%%%%%%%%%%%%%%%%
%%%%%%%%%%%%%%%%%%%%%%%%%%%%%%%%%%%%%%%%%%%%%%%%%%%%%%%%%%%%%%%%%%%%%%%%%%%%%%%%%%%%%
\section{Data and Methods} \label{data}

\subsection{Target Galaxies}

The nine massive \squiggle PSBs ($\log\,(\Mstar/\Msol) \approx 11.0-11.5$; \citealt{suess_mathrmsquiggecle_2022}) in this work have been selected from a larger sample of ALMA CO(2--1) measurements which consists of 50 objects \citep{bezanson_now_2022,setton_squigg_2025}. Figure \ref{fig:MgasSFR} shows the primary selection property for these nine galaxies, which were selected as the most molecular gas-rich systems (\MHt $>10^{10.25}$\,\Msol) ensuring efficient detection of the CO(5--4) transition. Two additional objects meeting this criterion were excluded because their redshifts place the CO(5--4) line in a region of poor atmospheric transmission in ALMA Band 7. \MHt quantities for the sample were derived assuming thermalized emission $r_{21}=1.0$ and the CO-to-H2 conversion factor $\alphaco = 4.0$ (\citealt{bezanson_now_2022,setton_squigg_2025}; see these works for further discussion on these assumptions). As with all molecular gas detections from \squiggle the nine galaxies included in this study possess reservoirs that are significantly offset from `normal' galaxies \citep{saintonge_cold_2011, tacconi_phibss_2013}, with gas masses up to nearly two orders of magnitude greater than expected given the low apparent SFRs. Likewise the selected objects are among the youngest of the \squiggle ALMA measurements, with the time since the onset of quenching ($t_q$, see \citealt{suess_recovering_2022} for more details on this definition) $\lesssim 200$\,Myr for the entire sample \citep{bezanson_now_2022,suess_mathrmsquiggecle_2022,setton_squigg_2025}. The CO luminosity fades on a decay timescale of $\approx$75-85\,Myr with depletion timescales of $\approx$1\,Gyr, suggesting that the rapid gas decline cannot be driven simply by consumption via residual star formation \citep{bezanson_now_2022,setton_squigg_2025}.  

\begin{deluxetable*}{ccccccccc}[ht!]
    \tablecaption{ALMA CO properties for nine gas-rich \squiggle PSBs. \label{tab:COProps}}
    \tablecolumns{8}
    \tablewidth{0pt}
    \setlength{\tabcolsep}{6.5pt}
    \tablehead{
        \colhead{Object} &
        \colhead{z} &
        \colhead{$L'_{\rm CO(2-1)}$\tablenotemark{a}} &
        \colhead{\MHt\tablenotemark{a}} &
        \colhead{$R_{\rm CO}$} &
        \colhead{$S_{\rm CO(5-4)}$} &
        \colhead{$L'_{\rm CO(5-4)}$} &
        \colhead{$L'_{\rm CO(5-4)}/L'_{\rm CO(2-1)}$} \\
        &
        &
        ($10^9$ K km s$^{-1}$ pc$^{2}$) &
        ($10^{10}$ \Msol) &
        (kpc) &
        (Jy km s$^{-1}$) &
        ($10^9$ K km s$^{-1}$ pc$^{2}$) &
    }
    \startdata
    J0907+0423 & 0.6635 & 14.41 $\pm$ 0.17 & 5.76 $\pm$ 0.07 & 1.79 $\pm$ 0.42 & 2.59 $\pm$ 0.23 & 2.57 $\pm$ 0.23 & 0.18 $\pm$ 0.02 \\
    J0909$-$0108 & 0.7021 & 6.98 $\pm$ 1.01 & 2.79 $\pm$ 0.40 & 2.28 $\pm$ 0.38 & 2.07 $\pm$ 0.16 & 2.31 $\pm$ 0.18 & 0.33 $\pm$ 0.06 \\
    J0912+1523 & 0.7473 & 8.47 $\pm$ 0.36 & 10.53 $\pm$ 0.02 & 6.20 $\pm$ 0.80\makebox[0pt][l]{\tablenotemark{b}} & $<$0.47 & $<$0.60 & $<$0.07 \\
    J1142+0006 & 0.5935 & 11.22 $\pm$ 0.24 & 4.49 $\pm$ 0.10 & 2.80 $\pm$ 0.56 & 4.45 $\pm$ 0.39 & 3.52 $\pm$ 0.31 & 0.31 $\pm$ 0.03 \\
    J1157+0132 & 0.7559 & 28.26 $\pm$ 0.40 & 11.30 $\pm$ 0.16 & 3.22 $\pm$ 0.36 & 6.54 $\pm$ 0.73 & 8.47 $\pm$ 0.94 & 0.30 $\pm$ 0.03 \\
    J1436+0447 & 0.6339 & 6.94 $\pm$ 0.42 & 2.78 $\pm$ 0.17 & 4.24 $\pm$ 0.51 & 0.89 $\pm$ 0.24 & 0.81 $\pm$ 0.22 & 0.12 $\pm$ 0.03 \\
    J1448+1010 & 0.6462 & 3.21 $\pm$ 0.21 & 1.28 $\pm$ 0.08 & 3.43 $\pm$ 0.39 & 1.93 $\pm$ 0.25 & 1.82 $\pm$ 0.24 & 0.57 $\pm$ 0.08 \\
    J2258+2313 & 0.7058 & 4.03 $\pm$ 0.23 & 1.61 $\pm$ 0.09 & 3.47 $\pm$ 0.77 & 0.77 $\pm$ 0.19 & 0.87 $\pm$ 0.21 & 0.22 $\pm$ 0.05 \\
    J2310$-$0047 & 0.7378 & 8.70 $\pm$ 0.34 & 3.48 $\pm$ 0.14 & 1.91 $\pm$ 0.59 & 1.38 $\pm$ 0.13 & 1.70 $\pm$ 0.16 & 0.20 $\pm$ 0.02 \\
    \enddata
    \tablenotetext{a}{J1448+1010 and J2258+2313 host extended gas reservoirs, so only the properties associated with the central galaxies are reported here \citep{spilker_star_2022,donofrio_quenching_2025}. The ALMA CO(2--1) properties for the rest of the sources are from \citet{bezanson_now_2022} and \citet{setton_squigg_2025}.}
    \tablenotetext{b}{The reported size for J0912+1523 corresponds to the CO(2--1) data \citep{bezanson_now_2022} since the object is a nondetection in the CO(5--4) data.}
    \tablecomments{Upper limits for nondetections are $3\sigma$ and assume a line width of the respective CO(2--1) transition. $R_{\rm CO}$ is the deconvolved circularized radius of the CO(5--4) emission. Line ratios are given in brightness temperature units.}
\end{deluxetable*}
\vspace{-0.85cm}

\subsection{SFR Measurements}
In this work we use two complementary SFR modeling frameworks for the \squiggle sample.
For the full \squiggle ALMA sample (see Figure \ref{fig:MgasSFR}), we use the conventional SFRs from the Spectral Energy Distribution (SED) fitting from \citet{suess_mathrmsquiggecle_2022}. These were derived from SDSS optical spectroscopy and optical-near-infrared photometry and correspond to the instantaneous ($\sim$10\,Myr averaged) SFR implied by best-fit nonparametric star formation histories (SFHs). The reported SFR corresponds to the recent level implied by the declining SFHs and does not include an additional, independent recent burst component. To assess the impact of potential dust-obscured star formation on the CO excitation (see Sections \ref{sec:GenCOEx} and \ref{sec:J1448}), we use the updated SED-modeling results from \citet{setton_squigg_2025}, which incorporate mid- and far-infrared constraints and explicitly brackets the allowed present-day SFR with two dust prescriptions. In the `tied birth cloud' model, the attenuation toward the youngest stars is linked to the diffuse ISM dust, again favoring a rapidly declining SFH without permitting a separate, heavily obscured final burst. In contrast the `free birth cloud' model introduces an independent final 10\,Myr bin which can allows for attenuation by optically thick birth cloud dust, allowing for a compact dust buried $\sim$10\,Myr burst that is weak in the rest-optical but luminous in the infrared. Throughout the rest of this work, we adopt the paired \citet{setton_squigg_2025} values as our SFR estimates when discussing obscured star formation activity. We refer the reader to these works for more specific details on the measurement of the respective SFRs.      

\subsection{ALMA CO Observations}

To investigate the physical state of the molecular gas in $z\sim0.7$ PSBs, we obtained ALMA Band 7 observations of the CO(5--4) transition ($\nu_{\mathrm{rest}}=576.27$\,GHz) for the nine \squiggle PSBs described above (PI: D'Onofrio; Program \#2024.1.01252.S). These observations complement the existing CO(2--1) data \citep{bezanson_now_2022,setton_squigg_2025}, enabling us to probe the CO excitation in these systems. Each target was observed with an average on-source integration time of $\approx$25\,minutes per source.

All data were calibrated with the standard ALMA pipeline in \texttt{CASA}. Continuum emission was detected in most cases and was subtracted in the $uv$ plane prior to imaging; further analysis on the continuum data will be presented in future work. Imaging was carried out using natural weighting to maximize sensitivity, consistent with previous analyses of \squiggle CO(2--1) data \citep{bezanson_now_2022,setton_squigg_2025}. The resulting spectral cubes have a spectral resolution of 50\,\kms and spectral line sensitivities that range $\approx$300-500\,\uJybeam per channel. The observations yield angular resolutions of $\approx$0.5\arc, corresponding to $\approx$3-4\,kpc at the redshifts of \squiggle galaxies.

For two galaxies J0912+1523 and J1448+1010 we also obtained ALMA Band 6 and 7 observations, respectively, of the CO(4--3) transition ($\nu_{\rm rest}=461.04$\,GHz) as ancillary objectives of previous ALMA programs (PI: Suess; Programs \#2018.1.01240.S, \#2018.1.01264.S) which we include in this work for completeness. The data reduction and imaging procedures for the CO(4--3) data follow the same methodology as the CO(5--4) observations. The final spectral cubes for both objects have a spectral resolution of 50\,\kms, a spectral line sensitivity of $\approx$150\,\uJybeam per channel, and an angular resolution of $\approx$1\arc.

%%%%%%%%%%%%%%%%%%%%%%%%%%%%%%%%%%%%%%%%%%%%%%%%%%%%%%%%%%%%%%%%%%%%%%%%%%%%%%%%%%%%%
%%%%%%%%%%%%%%%%%%%%%%%%%%%%%%%%% Results & Discussion %%%%%%%%%%%%%%%%%%%%%%%%%%%%%%
%%%%%%%%%%%%%%%%%%%%%%%%%%%%%%%%%%%%%%%%%%%%%%%%%%%%%%%%%%%%%%%%%%%%%%%%%%%%%%%%%%%%%
\section{Results and Discussion} \label{Results}

\subsection{CO(5--4) Emission Detected in Most Gas-rich \squiggle PSBs}

\begin{figure*}[ht!]
\centering
\hspace{0.0cm}
\includegraphics[width=0.95\textwidth]{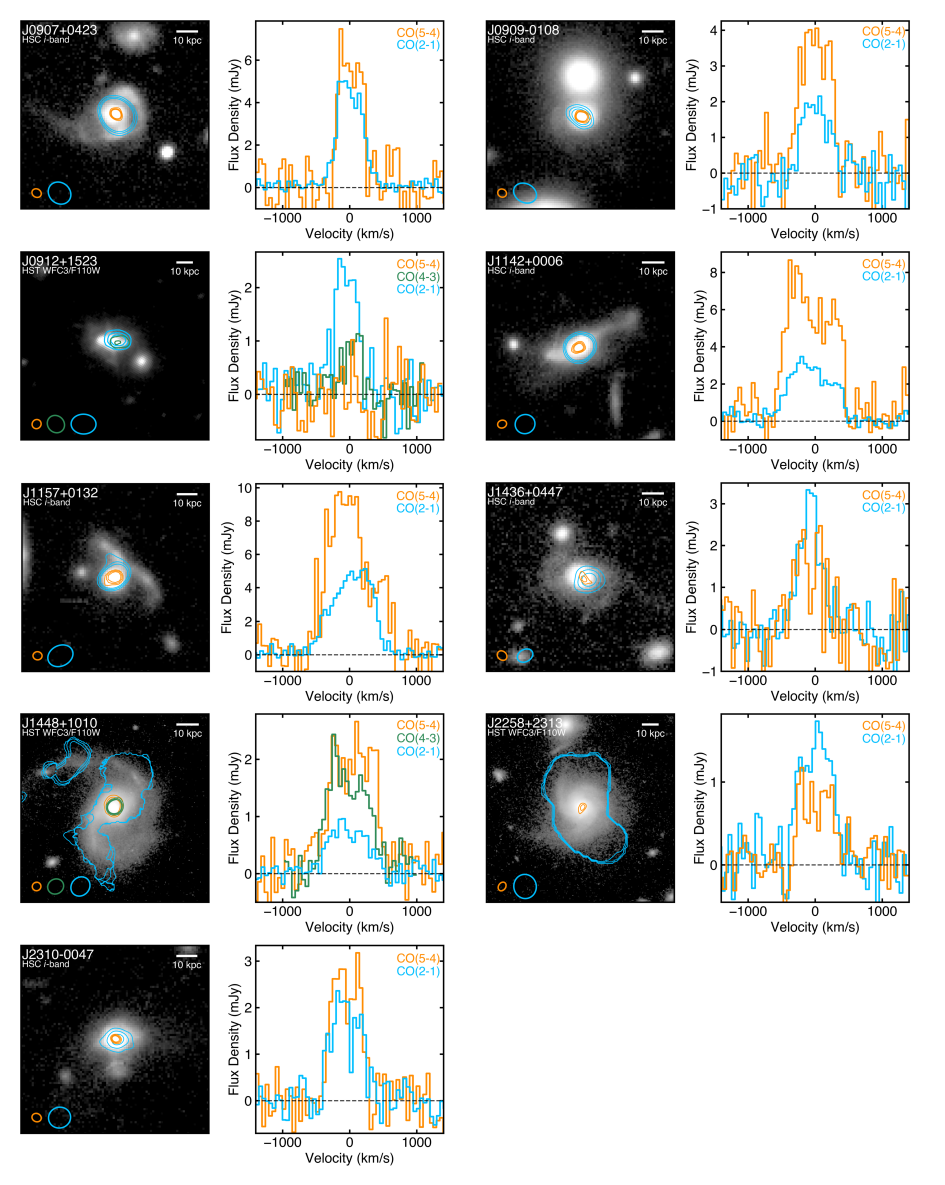}
\caption{The left panels show integrated emission contours of ALMA CO(2--1) (blue) and CO(5--4) (orange), and where available CO (4--3) (green), overlaid on best available optical cutouts ($14\arc \times 14\arc$) from HSC $i$-band \citep{aihara_hyper_2018} or HST WFC3/IR F110W \citep{suess_mathrmsquiggecle_2022,donofrio_quenching_2025,setton_squigg_2025}. Contours are shown at 3, 4, and 5$\sigma$ for each line, and the respective ALMA synthesized beams are indicated in the lower left corner. The right panels show the extracted spectra of the respective CO data cubes.    
}\label{fig:CO54gallery}
%\vspace{0.71cm}
\end{figure*}

Here, we analyze the CO(5-4) emission in nine gas-rich \squiggle PSBs. Figure \ref{fig:CO54gallery} presents a gallery of the CO(5--4) observations alongside existing CO(2--1) measurements \citep{bezanson_now_2022,setton_squigg_2025}. The left panel shows the best available optical imaging from Hyper Suprime-Cam (HSC) $i$-band \citep{aihara_hyper_2018} or Hubble Space Telescope (HST) Wide Field Camera 3/Infrared (WFC3/IR) F110W \citep{spilker_star_2022,suess_mathrmsquiggecle_2022,donofrio_quenching_2025} with overlaid contours of the CO(5--4) and CO(2--1) integrated emission, with the respective synthesized beam sizes shown in the lower left corner. We show in the right panels the corresponding CO spectra. We also include CO(4--3) spectra and contours of integrated emission maps where available.

The CO spectra were extracted using apertures defined from two-dimensional Gaussian fits to the CO(5--4) integrated emission maps using the \texttt{IMFIT} task in \texttt{CASA}. Rather than adopting the same aperture sizes used for the lower-resolution CO(2--1) data ($\sim$2\arc), this approach ensures that we isolate emission associated with the galaxy at the higher resolution of the CO(5--4) data. Integrated line fluxes were derived by fitting each spectrum with a single Gaussian profile, yielding values consistent with both direct channel summation and the fluxes returned via \texttt{IMFIT}. We produced tapered image cubes matched to the $\sim$1-2\arc beam size of the CO(2--1) data to test whether any extended CO(5--4) emission was resolved out. The tapered and natural-weighted images recover consistent integrated fluxes, indicating that our CO(5--4) measurements are not biased by resolution effects. For J0912+1523 we do not detect significant emission in the 50\,\kms spectral cube, so we additionally image spectral cubes binned to 100 and 150\,\kms channels. We do not detect $\geq3\sigma$ emission in either cube and classify this object as a nondetection, so a $3\sigma$ upper limit is computed assuming the line width of the CO(2--1) transition of the galaxy. The same procedures were applied to the two galaxies with ancillary CO(4--3) observations. 

We detect CO(5--4) emission in 8/9 galaxies, with $r_{52}=\ratio$ spanning a range of $\approx$0.1-0.3 (Table \ref{tab:COProps}) for most sources. J1448+1010 is a clear outlier with $r_{52}\approx0.6$, indicative of markedly higher excitation, therefore we treat further analysis of this object separate from the other sources. For all detections the CO(5--4) emission is resolved at the $\approx$0.5\arc resolution of the data. For the majority of sources, the emission for either CO transition is firmly identified as a single, central component and not associated with extended regions of the system, even in tapered images. However J1448+1010 and J2258+2313 exhibit extended CO(2--1) emission associated with tidal tails in their systems, so we use the central regions defined in their integrated CO(2--1) maps (see \citet{spilker_star_2022,donofrio_quenching_2025}) to measure $r_{52}$, since the CO(5-4) emission is only detected in the galaxy centers. For the two systems with CO(4--3) measurements, we find $r_{42}=0.11\pm0.05$ for J0912+1523 while \citet{donofrio_quenching_2025} reports $r_{42}=0.61 \pm 0.03$ for J1448+1010. Three galaxies in our sample overlap with those presented by \citet{zanella_gas_2025} (J0912+1523, J1448+1010, and J2258+2313), where the CO(5--4) flux measurements are generally consistent within the uncertainties. The only notable exception is J2258+2313 for which the reported flux in \citet{zanella_gas_2025} is approximately a factor of three higher than reported in this work, likely reflecting the lower signal-to-noise ratio of the data from the former study. Nonetheless, the range in CO excitation we find is similar to the ratios found in \citet{zanella_gas_2025} for eight $z\sim 0.6-1.3$ PSBs.    

\subsection{\squiggle PSBs Typically Host Moderate CO Excitation}
\label{sec:COSLED}

\begin{figure*}[ht!]
\centering
\hspace{0.0cm}
\includegraphics[width=0.999\textwidth]{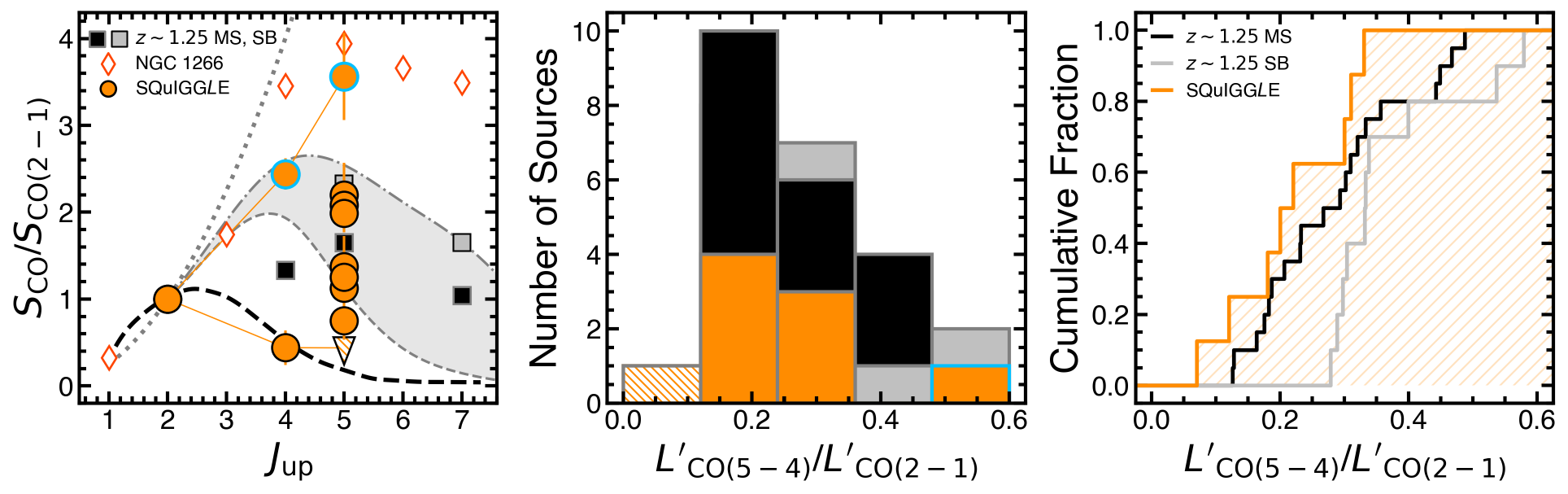}
\caption{CO SLEDs and excitation ratio distributions for the nine \squiggle PSBs. The left panel shows the CO SLEDs constructed from the CO(2--1) and CO(5--4) transitions (and CO(4--3) where available). The \squiggle PSBs are shown as orange circles (with the nondetection shown as an orange hatched triangle), with the AGN J1448+1010 outlined in blue. We compare to $z\sim1.25$ MS (black squares) and SB (gray squares) galaxies \citep{valentino_co_2020}, the Milky Way inner disk (black dashed line; \citealt{fixsen_cobe_1999}), the thermalized $S_{\rm CO}\propto J^2$ relation (dotted gray line), the model excitation sequence from \citet{narayanan_theory_2014} spanning $\Sigma_{\rm SFR}=0.1-10\,{\rm M_\odot\,yr^{-1}\,kpc^{-2}}$ (gray curve; dashed and dashed-dotted lines represent lower and upper bounds, respectively), and the nearby shocked PSB system NGC 1266 (red diamonds; \citealt{alatalo_discovery_2011,pellegrini_shock_2013}). The middle panel shows the distribution of \ratio for the \squiggle PSBs (with J1448+1010 outlined in blue) and the $z\sim1.25$ comparison galaxies, while the right panel shows the cumulative distributions for the same samples. The single nondetection is included in the histogram (orange hatched bin) and the cumulative distribution at the $3\sigma$ limit; however, the AGN J1448+1010 is not included in the latter. Overall, the sample is comprised of mostly moderate excitation CO SLEDs, with a single high excitation outlier in the AGN host J1448+1010.
}\label{fig:COSLED.pdf}
%\vspace{0.71cm}
\end{figure*}

The CO excitation of \squiggle PSBs provides a direct indication of whether the remaining molecular gas exists in a phase conducive to further star formation or in a low excitation state in these systems. We use the CO spectral line energy distribution (SLED) to characterize the physical state of the molecular gas, where the shape of the SLED reflects the relative contribution of cold, diffuse gas traced by low-$J$ transitions and warm, dense gas traced by higher-$J$ lines (e.g., \citealt{weis_highly-excited_2007,narayanan_theory_2014,kamenetzky_recovering_2018}). Figure \ref{fig:COSLED.pdf} presents the CO SLED for the nine PSBs. The left panel compares the \squiggle SLEDs with those of $z\sim1.25$ main-sequence (MS) and starburst (SB) galaxies \citep{valentino_co_2020}\footnote{This sample provides the most robust homogenous intermediate-redshift CO(5--4) reference, as no comparably uniform $z\sim0.7$ starburst and main-sequence CO(5--4) compilation is currently available.}, the Milky Way inner disk \citep{fixsen_cobe_1999}, the thermalized $S_{\rm CO}\propto J^2$ relation, and the star formation driven model excitation sequence from \citet{narayanan_theory_2014} spanning star formation surface densities of $\Sigma_{\rm SFR}=0.1-10\,{\rm M_\odot\,yr^{-1}\,kpc^{-2}}$. We also include NGC 1266, a nearby PSB system exhibiting a compact, shocked molecular outflow \citep{alatalo_discovery_2011,pellegrini_shock_2013}. The middle panel shows the distribution of \ratio for the \squiggle sample compared to the $z\sim1.25$ comparison galaxies, while the right panel presents their cumulative distributions.

Taken together, the \squiggle PSBs show predominantly moderate ({$r_{52}\approx{0.1-0.3}$}) CO excitation in their molecular gas reservoirs. Seven detections along with the single nondetection show excitation comparable to the low end of the $z\sim1.25$ MS sample, and in some cases approaching the Milky Way inner disk. J1448+1010 is an outlier with an elevated excitation ratio comparable to the shocked PSB NGC 1266. This range is reflected in the middle and right panels of Figure \ref{fig:COSLED.pdf}, where most of the \squiggle objects are concentrated near the low end of the MS distribution of the $z\sim1.25$ sample. A two-sample Kolmogorov-Smirnov test of the \squiggle galaxies including the nondetection but excluding J1448+1010 yields $p$-values of 0.607 and 0.037 when compared to the MS and SB populations, respectively. Although the sample size is small, this indicates that we can reject consistency with the $z\sim1.25$ SB distribution at the 2.1$\sigma$ level, but cannot reject consistency with the MS distribution. 

\begin{figure*}[ht!]
\centering
\hspace{0.0cm}
\includegraphics[width=0.95\textwidth]{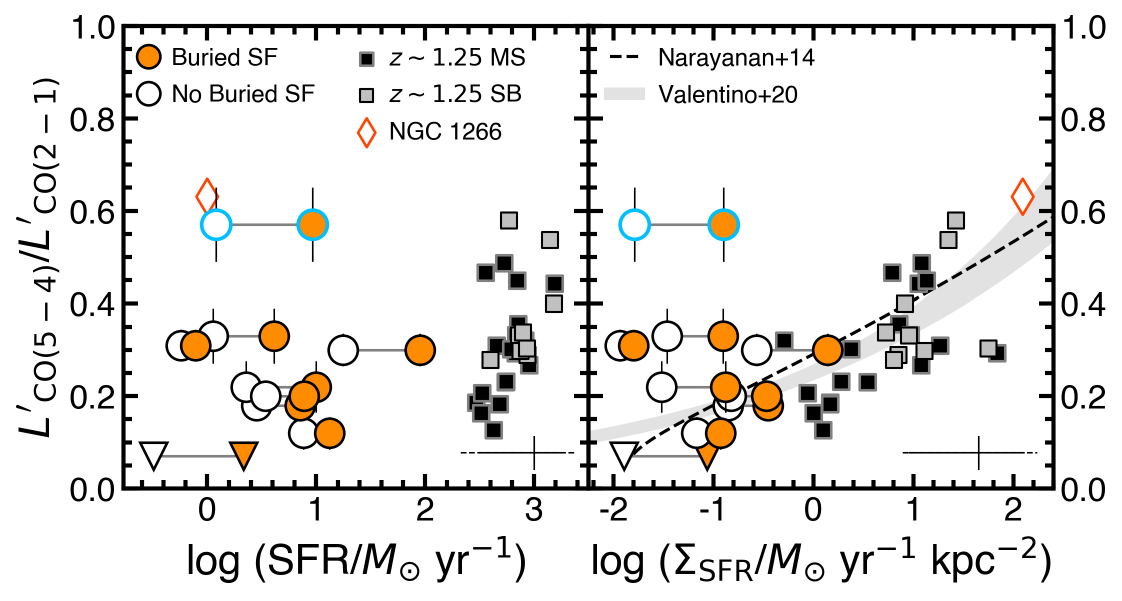}
\caption{The left panel shows \ratio against total SFR, showing both the non-buried (white circles) and buried (orange circles) SFR estimates \citep{setton_squigg_2025}. In the right panel we show the excitation ratio versus $\Sigma_{\rm SFR}$ computed using CO(5--4) source sizes (see Table \ref{tab:COProps}). As in Figure \ref{fig:COSLED.pdf}, the AGN J1448+1010 is outlined in blue. Both panels include $z\sim1.25$ MS (gray squares) and SB (black squares) comparison galaxies from \citet{valentino_co_2020} and the shocked PSB NGC 1266 (red diamond; \citealt{alatalo_discovery_2011,pellegrini_shock_2013,alatalo_escape_2015}) for reference. The right panel additionally shows the observed trend from \citet{valentino_co_2020} and the excitation model from \citet{narayanan_theory_2014}. Black crosses indicate median uncertainties for the \squiggle objects; for the SFR and $\Sigma_{\rm SFR}$ axes, solid and dashed bars indicate those for the buried and non-buried values, respectively. Together, no clear trend is observed between the CO excitation and both total or surface-density SFR for \squiggle PSBs, where instead all systems are consistent with little to modest star-forming activity.
}\label{fig:r52SFR}
%\vspace{0.71cm}
\end{figure*}

Elevated excitation ratios (upwards of $r_{52}\approx0.6$) generally trace gas that is both dense and warm, conditions typical of luminous infrared galaxies hosting significant levels of star formation or systems in which feedback and dynamical interactions strongly perturb the ISM (e.g., \citealt{papadopoulos_molecular_2012,pellegrini_shock_2013,spilker_rest-frame_2014, valentino_co_2020,birkin_almanoema_2021}). In contrast, low-excitation CO SLEDs correspond to diffuse, subthermally excited gas that is inefficient at forming stars (e.g., \citealt{crocker_atlas3d_2012,bayet_atlas3d_2013}), similar to that observed in low-redshift PSBs where large molecular reservoirs remain but with low dense gas fractions \citep{french_state_2023}. Most of the \squiggle PSBs have CO excitation ratios that are consistent with the low end of the $z\sim1.25$ MS distribution, where their cumulative distribution is generally shifted to ratios lower than the MS sample (Figure \ref{fig:COSLED.pdf}). This similarity is limited to the excitation and does not imply that the \squiggle PSBs are MS-like in their overall star-forming characteristics. The MS comparison galaxies have substantially higher SFR properties (see Section \ref{sec:GenCOEx}; Figure \ref{fig:r52SFR}) with stellar masses $\approx$0.5\,dex lower than the gas-rich \squiggle PSBs \citep{muzzin_discovery_2013,laigle_cosmos2015_2016}, likely uniquely affecting ISM conditions that contribute to the mid-$J$ CO emission. Additionally, they populate the CO SLED predicted by \citet{narayanan_theory_2014} for $\Sigma_{\rm SFR}\approx0.1-10\,{\rm M_\odot\,yr^{-1}\,kpc^{-2}}$, where most are consistent with the low end which suggests little to moderate ongoing star formation in compact central gas. J1448+1010 is the lone exception with excitation well above this regime, suggesting that an additional, non-stellar heating source is required to drive up the excited gas. Generally, however, the moderate excitation found in \squiggle PSBs suggests that the molecular gas is in an intermediate state between diffuse reservoirs and starburst conditions.  

\subsection{What is Generally Driving the CO Excitation in \squiggle PSBs?}
\label{sec:GenCOEx}

Understanding what influences the CO excitation of the molecular gas in PSBs is important in identifying the processes that regulate its physical state. One possibility is that some \squiggle PSBs host ongoing, heavily dust-obscured star formation, which can elevate gas temperatures and densities through heating and thereby enhance mid-$J$ CO excitation (e.g., \citealt{papadopoulos_molecular_2012,valentino_co_2020}). \citet{setton_squigg_2025} showed that allowing for an obscured star-forming component can increase the inferred SFRs of gas-rich \squiggle PSBs by up to $\sim$0.5 dex. If such obscured star formation is present, galaxies with higher inferred buried SFRs should preferentially exhibit higher CO excitation. While this test alone cannot definitively rule out the buried star formation scenario, it provides a direct way to assess whether ongoing star formation is plausibly driving the observed excitation. Additionally, previous studies have shown that CO excitation correlates more strongly with $\Sigma_{\rm SFR}$ than with total SFR or infrared luminosity, reflecting the sensitivity of CO excitation to local ISM conditions (e.g., \citealt{narayanan_theory_2014,valentino_co_2020}). We therefore examine how the excitation ratios of the \squiggle PSBs relate to both total SFR and $\Sigma_{\rm SFR}$ to assess whether the excitation conditions reflect potentially ongoing obscured star formation.

Figure \ref{fig:r52SFR} compares the excitation ratios of the nine \squiggle PSBs with their star formation activity. The left panel shows \ratio against the total SFR, where each galaxy is plotted with both the non-buried and buried SFR estimates from the spectral energy distribution modeling including mid- and far-infrared photometry in \citet{setton_squigg_2025}; see \citet{setton_squigg_2025} for further details on the SFR estimates. Comparison samples of $z\sim1.25$ galaxies from \citet{valentino_co_2020} are shown for reference, along with NGC 1266 for which we show the scenario of a nuclear starburst, the region where the highly excited gas is hosted (see \citealt{alatalo_escape_2015} for further details). The right panel presents the same ratio as a function of $\Sigma_{\rm SFR}$, computed using the deconvolved circularized radii of the CO(5--4) sources listed in Table \ref{tab:COProps}. The use of CO(5--4) sizes provide a more appropriate estimate of $\Sigma_{\rm SFR}$, as these data offer higher spatial resolution than the CO(2--1) observations and preferentially trace the dense, possibly star-forming molecular gas. We determined the spatial extent of the CO emission by fitting two-dimensional Gaussian profiles to the $\approx$0.5\arc angular resolution CO(5--4) integrated intensity maps using the \texttt{IMFIT} task in \texttt{CASA} to obtain deconvolved source sizes. For the source (J0912+1523) not detected in CO(5--4), we use the deconvolved size reported in \citet{bezanson_now_2022} from the $\approx$1.5\arc beam size CO(2--1) data. The observed trend from \citet{valentino_co_2020} spanning local spirals to various high-redshift populations, as well as model predictions from \citet{narayanan_theory_2014} are also included for comparison. 

The \squiggle galaxies show no clear trend between excitation ratio and either total or surface-density SFR. In the left panel of Figure \ref{fig:r52SFR} systems with the highest excitation do not exhibit the highest buried SFRs, though most of the sample only spans a relatively limited excitation range. This is consistent with the conclusions of \citet{setton_squigg_2025}, which found no strong preference for the model invoking substantial obscured star formation over those in which star formation is genuinely suppressed. As noted in \citet{setton_squigg_2025} only J1157+0132 possesses an obscured SFR estimate that is consistent with an active starburst, while the others remain relatively modest even under the maximally buried model, consistent with modest CO excitation of $r_{52}\approx0.1-0.3$.

In the $\Sigma_{\rm SFR}$ plane (right) the PSBs occupy the low end of the empirical and theoretical relations particularly for the buried SFRs, with $\Sigma_{\rm SFR}\sim0.01-1$\,M$_\odot$\,yr$^{-1}$\,kpc$^{-2}$ when considering both fits, consistent with little to moderate ongoing star formation. It is important to note that the $\Sigma_{\rm SFR}$ values predicted from their CO SLEDs are offset an order of magnitude with $\Sigma_{\rm SFR}\sim0.1-10$\,M$_\odot$\,yr$^{-1}$\,kpc$^{-2}$ (see Section \ref{sec:COSLED}, Figure \ref{fig:COSLED.pdf}). This offset implies that their excitation is decoupled from present-day star formation (obscured or not), but still within a regime where little to modest star formation in compact regions can reproduce the observed excitation. Modest decoupling is expected given that the buried fits have instantaneous (averaged over the last 10\,Myr) SFRs while the CO-luminous sample likely reached peak SFRs of order a few 100\,\sfr within the past few 100\,Myr \citep{setton_squigg_2025}, thus the CO SLED likely does not track exactly the current SFR in this post-merger, post-burst phase. Overall, the observed excitation is consistent with being reproduced by little to moderate ongoing star formation confined to compact regions. In this picture, the gas-rich \squiggle PSBs likely reflect a post-merger, post-burst state defined by relatively suppressed star formation efficiencies rather than strong buried starburst activity.   

Possible non-stellar heating cannot be ruled out, but current indicators are not conclusive. Based on [O III]/H$\beta$ line ratios only two systems (J1436+0447 and J1157+0132) show optical AGN signatures \citep{greene_role_2020}, yet their SED models do not require a significant mid-infrared AGN component \citep{setton_squigg_2025}. J2258+2313 shows evidence for an elevated central gas dispersion while the origin of its radio continuum emission (VLA 6\,GHz) remains inconclusive, likely indicative of turbulence driven by a merger evidenced by the $\sim$50\,kpc long tidal tails present in the system \citep{donofrio_quenching_2025}. However its mid-$J$ excitation likewise remains relatively modest, suggesting that merger-driven turbulence alone is perhaps insufficient in driving high CO excitation in these systems. Merger-driven turbulence in the local universe has been shown to drive up gas excitation, although typically accompanied by a starburst (e.g., \citealt{papadopoulos_molecular_2012}). We note that J0912+1523 shows no obvious AGN indicators and as the oldest system in this sample has limited ongoing star formation even when allowing for obscured activity \citep{setton_squigg_2025}, possibly explaining the undetected CO(5--4) emission for this source. Future analysis of in-hand Very Large Array observations complementary to the full ALMA sample presented in \citet{bezanson_now_2022} and \citet{setton_squigg_2025} will better isolate radio-mode feedback properties and test for systematic links with radio emission and quenching.

While we find that little to modest current star formation can contribute to the observed CO excitation levels, this does not preclude contribution from further star formation activity in some systems. As argued by \citet{setton_squigg_2025} a plausible pathway is the rejuvenation of star formation, in which the molecular gas persists in a suppressed, low-efficiency phase before fueling a later burst. This is qualitatively supported by the fact that all of the galaxies presented in this work exhibit clear tidal features \citep{spilker_star_2022,verrico_merger_2023,donofrio_quenching_2025}, suggesting relatively recent mergers, where in low-redshift systems elevated PSB fractions occur upwards of $\sim$500\,Myr after the merger \citep{ellison_galaxy_2024,ellison_galaxy_2025}. This picture is consistent with simulations of $z\sim0.7$ PSBs which exhibit lulls punctuated by intermittent star formation \citep{cenci_nature_2025}. Likewise, Savage \etal (2026, in preparation) indicates that periodic AGN jet power can depress central gas densities and stall star formation while leaving substantial reservoirs available for subsequent activity. Thus, it cannot be ruled out that the combination of moderate excitation and low current non-buried SFRs in most of the gas-rich \squiggle PSBs may instead reflect the temporary stalling of star formation rather than permanent quiescence.

\subsection{What is Contributing to the High Excitation in J1448+1010?}
\label{sec:J1448}

Most of the gas-rich \squiggle PSBs exhibit CO excitation consistent with little to moderate ongoing star formation. However, the outlier galaxy J1448+1010 is the single gas-rich \squiggle PSB with mid-$J$ CO emission beyond what is reasonably expected from solely current star formation, including the buried SFR estimate. Reproducing such CO excitation ($r_{52}\approx0.6$) from star formation alone is typically found in high-redshift submillimeter systems with SFR $\geq100\,\sfr$ (e.g., \citealt{spilker_rest-frame_2014,birkin_almanoema_2021}). This excess could also arise from the combination of AGN activity and/or merger-driven turbulence with any potential ongoing star formation. J1448+1010 resembles J2258+2313 in its high central gas dispersion and extended tidal tails, suggestive of merger-driven turbulence \citep{donofrio_quenching_2025}. However, J1448+1010 also shows clear signatures of AGN activity. Based on the [O III]/H$\beta$ ratio J1448+1010 shows optical AGN activity \citep{greene_role_2020}, while also having a significant mid-IR AGN contribution \citep{setton_squigg_2025}. Additionally, the VLA 6\,GHz radio continuum indicates this galaxy hosts clear compact radio jets \citep{donofrio_quenching_2025}. Considering that the extended tidal features (upwards of $\sim$60\,kpc in length) in this system show low gas dispersions and some localized star formation activity \citep{donofrio_quenching_2025}, the energy from the AGN is likely not input into the outskirts but instead driving up the excitation in the ISM. Similar AGN-induced turbulence and shock heating have been inferred in nearby systems, where interactions with the ISM drive localized increases in CO excitation and suppress star formation in circumnuclear gas (e.g., \citealt{pellegrini_shock_2013,garcia-burillo_molecular_2014,alatalo_escape_2015}). For J1448+1010 the SED fits that allow for a buried component yield SFRs about an order of magnitude higher than the non-buried solutions, which could possibly add a compact star-forming contribution to the excitation. Taken together, this suggests AGN-driven energy input as a significant driver of the CO excitation, supplemented by modest obscured star formation and/or merger-driven turbulence.

A similar object at low-redshift is the shocked PSB NGC 1266, which hosts a very compact ($\leq$100\,pc) nuclear molecular region with a CO ladder similar to J1448+1010 (\citealt{alatalo_discovery_2011,pellegrini_shock_2013}; Figure \ref{fig:COSLED.pdf}). In this system a modest nuclear star-forming component and AGN activity are both likely to contribute to the central excitation of NGC 1266, with shocks required to power the mid- to high-$J$ CO \citep{alatalo_discovery_2011,pellegrini_shock_2013,otter_pulling_2024}. J1448+1010 presents a similar scenario, where the elevated CO(5-4) emission is centrally concentrated with the combination of clear AGN activity and possible low-level hidden star formation contributing to the observed excitation. These parallels suggest that J1448+1010 may be a qualitative higher-redshift analog of the shocked PSB NGC 1266. However, in the case of buried star-forming activity $\Sigma_{\rm SFR}$ is still multiple orders of magnitude lower for J1448+1010 than NGC 1266 (Figure \ref{fig:r52SFR}). This indicates that perhaps a radiatively powerful AGN is necessary for the observed excitation level in J1448+1010. Overall, this places J1448+1010 as a PSB distinct from the rest of the \squiggle gas-rich sample, caught in a different stage post-merger with clear AGN activity.  

%%%%%%%%%%%%%%%%%%%%%%%%%%%%%%%%%%%%%%%%%%%%%%%%%%%%%%%%%%%%%%%%%%%%%%%%%%%%%%%%%%%%%
%%%%%%%%%%%%%%%%%%%%%%%%%%%%%%%%% Conclusions %%%%%%%%%%%%%%%%%%%%%%%%%%%%%%%%%%%%%%%
%%%%%%%%%%%%%%%%%%%%%%%%%%%%%%%%%%%%%%%%%%%%%%%%%%%%%%%%%%%%%%%%%%%%%%%%%%%%%%%%%%%%%
\section{Conclusions} \label{Conclusions}

In this work we present ALMA CO(5--4) observations of nine gas-rich PSBs at $z\sim0.7$ from the \squiggle survey, providing a direct view of the physical state of the molecular gas during the onset of quenching. We find that most systems exhibit moderately excited CO SLEDs comparable to those found in some $z\sim1.25$ MS galaxies, while J1448+1010 is a clear outlier that likely requires non-stellar heating in addition to any compact, obscured star formation. We do not find a clear trend between CO excitation and either total or surface-density SFR, though the sample is small and spans a limited dynamic range in excitation, where additional mid-$J$ CO observations of the less CO-luminous \squiggle PSBs could help extend the $r_{52}$ plane. The mid-$J$ CO data alone do not demand large hidden SFRs, but importantly they also presently cannot rule out obscured activity.

The detection of moderately excited CO in most of the sample sets the physical state of the gas reservoirs in these \squiggle PSBs apart from their low-redshift counterparts. Lower redshift PSBs typically exhibit diffuse, low-excitation (albeit with low-$J$ CO ladders) molecular reservoirs with little evidence for the dense gas phases required for any star formation \citep{french_state_2023}. In contrast, the presence of moderately excited gas in \squiggle galaxies indicates that the quenching process may proceed differently earlier in the universe, where galaxies are more gas rich and have experienced more frequent gas-rich mergers. Molecular gas content increases with redshift as $(1+z)^{\approx2.5-3}$ (e.g., \citealt{tacconi_evolution_2020}), while the galaxy merger rate rises as $(1+z)^{\approx2-3}$ (e.g., \citealt{lotz_evolution_2008,conselice_structures_2009,bridge_cfhtls-deep_2010,duncan_observational_2019,ferreira_galaxy_2020}). An exception in the gas-rich \squiggle PSB sample is J1448+1010, with a relatively high excitation CO SLED and clear AGN signatures which closely resemble the nearby shocked PSB NGC 1266, where compact star formation and AGN-driven turbulence power the higher-$J$ CO emission. Together, the greater gas availability and more frequent interactions at $z\sim0.7$ may lead to denser, more turbulent ISM conditions that produce moderate CO excitation or greater in post-burst, post-merger systems. However, surface-density SFRs indicate that \squiggle PSBs are defined by relatively suppressed star formation efficiencies, consistent with previous \squiggle results \citep{bezanson_now_2022,setton_squigg_2025}.

Despite the CO(5--4) emission observed in the ISM of most of the sample, we find no evidence for the emission in the outskirts of any system, including J1448+1010 which has tidal tails observed in both molecular gas traced by CO(2--1) observations and stellar continuum. The northern tail likely hosts ongoing star formation at an approximately MS rate, however only weak CO(4--3) emission ($r_{42}\sim0.1$) is detected \citep{donofrio_quenching_2025}. Deeper, lower-resolution observations sensitive to the low surface brightness emission of the outskirts will be necessary to probe the CO excitation of molecular gas in the outer regions of these systems. This is important considering that at least two of the \squiggle gas detections exhibit quenching driven by the tidal removal of cold gas \citep{spilker_star_2022,donofrio_quenching_2025}, and with $\sim$70\% of the youngest PSBs in the \squiggle sample exhibiting merger features \citep{verrico_merger_2023}, this quenching mechanism could be especially relevant at higher redshifts.     

Observations targeting additional CO transitions and dense gas tracers (e.g., HCN, HCO$^+$; \citealt{french_state_2023,lin_almaquest_2024}) will be important for disentangling the relative contributions of diffuse and dense molecular phases in these systems. Expanding the CO ladder will enable detailed radiative transfer modeling of the excitation conditions providing direct constraints on gas densities and temperatures, as seen in the analysis of low-redshift PSBs with multiple CO transitions \citep{french_state_2023}. However, variations in the CO-to-H$_2$ conversion factor $\alpha_{\rm CO}$ could also influence the CO properties and inferred gas masses (see discussions in \citealt{donofrio_quenching_2025} and \citealt{setton_squigg_2025}). Observations of CO isotopologues (e.g., $^{13}$CO; \citealt{teng_physical_2023}) and analysis of the dust continuum will further constrain $\alpha_{\rm CO}$ by probing optical depth and gas-to-dust ratio variations. These diagnostics can reveal whether differences in excitation among \squiggle PSBs indeed arise from variations in gas density and temperature, or perhaps from changes in \alphaco. Furthermore, complementary near-infrared spectroscopy of SFR tracers such as Pa$\alpha$ will refine obscured SFR estimates and test for residual star formation. Extending such analyses to a larger sample spanning a broader range in characteristics such as the molecular gas content and the time since the onset of quenching will establish a homogeneous view of the physical state of the ISM in higher-redshift PSBs.   
 
\section*{Acknowledgements}
V.R.D., J.S.S., and K.A.S. gratefully acknowledge support from the National Science Foundation under NSF-AAG No. 2407954 \& 2407955. V.R.D. also acknowledges support provided by the NSF through award SOSPA 11-006 from the NRAO.

This paper makes use of the following ALMA data: ADS/JAO.ALMA\#2016.1.01126.S, ADS/JAO.ALMA\#2017.1.01109.S, ADS/JAO.ALMA\\\#2018.1.01240.S, ADS/JAO.ALMA\#2018.1.01264.S, ADS/JAO.ALMA\#2019.1.00221.S, ADS/JAO.ALMA\\\#2021.1.00761.S, ADS/JAO.ALMA\#2021.1.00988.S, ADS/JAO.ALMA\#2021.1.01535.S, ADS/JAO.ALMA\\\#2024.1.01252.S. ALMA is a partnership of ESO (representing its member states), NSF (USA) and NINS (Japan), together with NRC (Canada), MOST and ASIAA (Taiwan), and KASI (Republic of Korea), in cooperation with the Republic of Chile. The Joint ALMA Observatory is operated by ESO, AUI/NRAO and NAOJ. The National Radio Astronomy Observatory is a facility of the National Science Foundation operated under cooperative agreement by Associated Universities, Inc. 

This research is based on observations made with the NASA/ESA HST obtained from the Space Telescope Science Institute, which is operated by the Association of Universities for Research in Astronomy, Inc., under NASA contract NAS 5–26555. These observations are associated with programs 16201, 16248, and 15436. All of the data presented in this paper were obtained from the Mikulski Archive for Space Telescopes (MAST) at the Space Telescope Science Institute. The specific observations analyzed can be accessed via \dataset[https://doi.org/10.17909/9xpd-3752]{https://doi.org/10.17909/9xpd-3752} and \dataset[https://doi.org/10.17909/p5bd-nn41]{https://doi.org/10.17909/p5bd-nn41}. Support to MAST for these data is provided by the NASA Office of Space Science via grant NAG5–7584 and by other grants and contracts.

The Hyper Suprime-Cam (HSC) collaboration includes the astronomical communities of Japan and Taiwan, and Princeton University. The HSC instrumentation and software were developed by the National Astronomical Observatory of Japan (NAOJ), the Kavli Institute for the Physics and Mathematics of the Universe (Kavli IPMU), the University of Tokyo, the High Energy Accelerator Research Organization (KEK), the Academia Sinica Institute for Astronomy and Astrophysics in Taiwan (ASIAA), and Princeton University. Funding was contributed by the FIRST program from the Japanese Cabinet Office, the Ministry of Education, Culture, Sports, Science and Technology (MEXT), the Japan Society for the Promotion of Science (JSPS), Japan Science and Technology Agency (JST), the Toray Science Foundation, NAOJ, Kavli IPMU, KEK, ASIAA, and Princeton University.

Based in part on data collected at the Subaru Telescope and retrieved from the HSC data archive system, which is operated by the Subaru Telescope and Astronomy Data Center at the National Astronomical Observatory of Japan.

This research has made use of NASA's Astrophysics Data System.

\facility{ALMA, HST (WFC3), SDSS, Subaru}

\software{
\texttt{astropy} \citep{astropy_collaboration_astropy_2013,astropy_collaboration_astropy_2018,astropy_collaboration_astropy_2022},
CASA \citep{casa_team_casa_2022},
\texttt{matplotlib} \citep{hunter_matplotlib_2007}}

%\clearpage
\bibliographystyle{aasjournal}
\bibliography{SquiggleGasExcitation.bib}

\end{CJK*}
\end{document}